\documentclass[sigconf]{acmart}
\usepackage{graphicx,color}
\usepackage{xcolor}
\usepackage{textcase}
\usepackage{setspace} 
\usepackage[linesnumbered,ruled,vlined]{algorithm2e}
\usepackage{epstopdf}
\usepackage{epsfig}
\usepackage{amsmath,amsthm}
\usepackage{multirow}
\usepackage{hhline}
\usepackage{verbatim}
\usepackage{algorithmicx}
\usepackage{lipsum}
\usepackage{algpseudocode}
\usepackage{cases}
\usepackage{enumitem}
\usepackage[tablename=TABLE]{caption}
\usepackage{hyperref}
\usepackage{setspace}
\usepackage{graphics}
\usepackage{graphicx}
\usepackage{subfig}
\usepackage{xspace}
\usepackage{textcomp}

\copyrightyear{2023}
\acmYear{2023}
\setcopyright{acmlicensed}\acmConference[ACM  MobiWac '23, ACM MSWIM '23]{Proceedings of the Int'l ACM Symposium on Mobility Management and Wireless Access}{October 30-November 3 2023}{Montreal, QC, Canada}
\acmBooktitle{Proceedings of the Int'l ACM Symposium on Mobility Management and Wireless Access (MobiWac '23), October 30-November 3 2023, Montreal, QC, Canada}

\newcommand{\fakepar}[1]{\vspace{.5mm}\noindent\textbf{#1.}}
\newcommand\figref[1]{Fig.\,\ref{#1}}
\newcommand\secref[1]{Sec.\,\ref{#1}}

\DeclareCaptionTextFormat{up}{\MakeTextUppercase{#1}}

\newcommand{\capt}[1]{\mdseries{\emph{#1}}}

\newcommand\eam{\textsc{EAM}\xspace}

\begin{document}
\title{Application-aware Energy Attack Mitigation\\ in the Battery-less Internet of Things\vspace*{-0.12in}}

\author[Singhal et al.]{Chetna Singhal$^{+}$, Thiemo Voigt$^{\dagger\ddagger}$, and Luca Mottola$^{*\dagger\ddagger}$}
\affiliation{\institution{$^+$Indian Institute of Technology (IIT) Kharagpur, $^*$Politecnico di Milano (Italy),\\ $^\dagger$RI.SE Sweden, $^\ddagger$Uppsala University (Sweden)}\country{}}
\begin{abstract}

We study how to mitigate the effects of energy attacks in the battery-less Internet of Things~(IoT).
Battery-less IoT devices \emph{live and die} with ambient energy, as they use energy harvesting to power their operation.
They are employed in a multitude of applications, including safety-critical ones such as biomedical implants.
Due to scarce energy intakes and limited energy buffers, their executions become \emph{intermittent}, alternating periods of active operation with periods of recharging their energy buffers.
Experimental evidence exists that shows how controlling ambient energy allows an attacker to steer a device execution in unintended ways: energy provisioning effectively becomes \emph{an attack vector}. 
We design, implement, and evaluate a mitigation system for energy attacks.
By taking into account the specific application requirements and the output of an attack detection module, we tune task execution rates and optimize energy management.   
This ensures continued application execution in the event of an energy attack.
When a device is under attack, our solution ensures the execution of 23.3\% additional application cycles compared to the baselines we consider and increases task schedulability by at least 21\%, while enabling a 34\%  higher peripheral availability.

\end{abstract}



\maketitle
\setcitestyle{numbers,sort,compress}

\section{Introduction}\label{sec:introduction}

Ambient energy harvesting allows Internet of Things~(IoT) devices to eliminate their dependency on traditional batteries~\cite{harvesting-survey}.
This enables drastic reductions of maintenance costs and previously unattainable deployments, for example, due to the reduction of a device's footprint.
Several battery-less IoT deployments exist, even in safety-critical settings such as biomedical implants~\cite{water-deployment-microbial-fuel-cell,tethys,soil-termoelectric,sensys20deployment,denby2023kodan, ea_base} and using a variety of energy sources~\cite{harvesting-survey}.

Energy harvested from the environment, however, is highly variable in time~\cite{harvesting-survey}, yet energy buffers, such as capacitors, need to be miniaturized as well, as they often represent a dominating factor in size.
This trait clashes with the push to realize tiny devices enabling pervasive deployments.
System shutdowns due to energy depletion are thus difficult to avoid and executions becomes \emph{intermittent}~\cite{awesome}: periods of active execution and periods of energy harvesting come to be unpredictably interleaved.

\begin{figure}[tb]
  \centering
    \includegraphics[width=.99\linewidth]{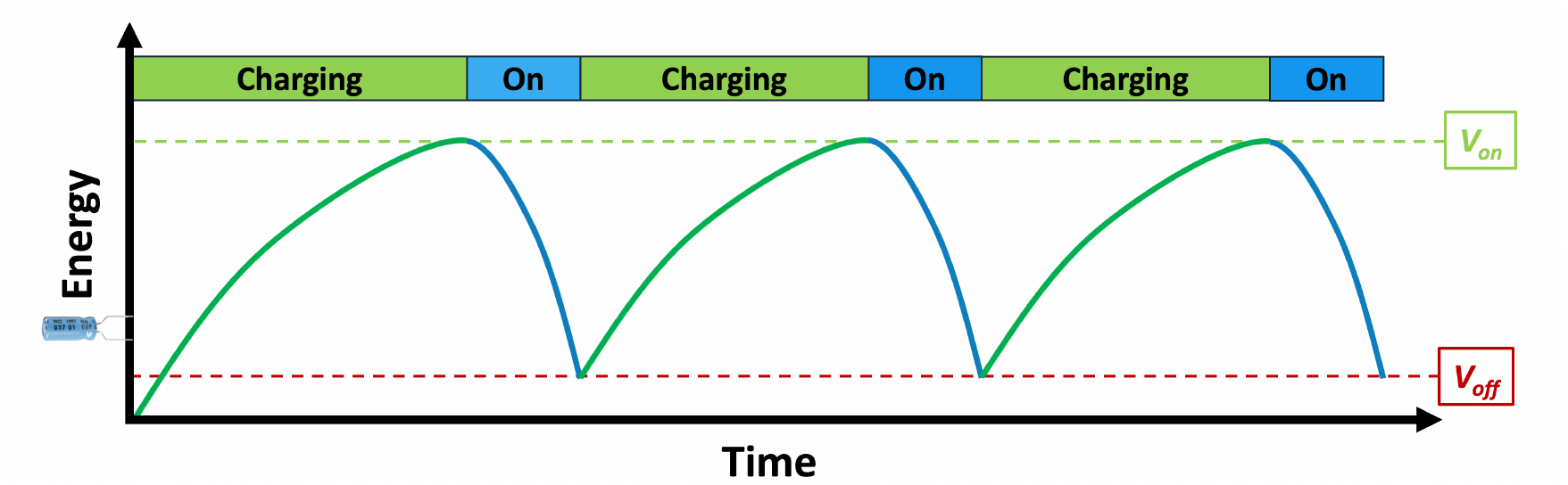}
    \vspace{-0.15in}
   \caption{Example intermittent execution. \capt{Periods of active computation and periods of charging the energy buffer alternate.}}
   \vspace{-0.2in}
   \label{fig:execution}
 \end{figure}

\fakepar{Computing intermittently} \figref{fig:execution} shows an example execution.
The ambient charges the onboard capacitor until a voltage threshold $V_{\mathit{on}}$ is reached that causes the device to power on.
The device starts sensing, computing, and communicating as long as the capacitor charge remains above a threshold $V_{\mathit{off}}$.

Using existing energy harvesting techniques and resource-con\-strain\-ed IoT platforms, such as TI MSP430 MCUs, the net energy balance during active execution is systematically negative, so the device rapidly reaches~$V_{\mathit{off}}$. 
When this happens, the device switches off, waiting for the capacitor to reach again the power-on threshold~$V_{\mathit{on}}$.
This intermittent execution may occur on tiny time scales; computing simple error correction codes on a battery-less IoT device powered through radio-frequency energy harvesting may require as many as 16 energy cycles~\cite{HarvOS}.

Using resource-constrained IoT devices, applications run on bare hardware with \emph{no operating system support}~\cite{awesome}.
When the device powers off upon reaching  $V_{\mathit{off}}$, the system state would normally be lost and the device would completely reboot the next time it reaches $V_{\mathit{on}}$.
To address this issue, intermittently-computing IoT systems employ ad-hoc techniques~\cite{HarvOS,mementos,Hibernus,Hibernus++,chinchilla,QuickRecall,Samoyed,ratchet,dice,alpaca,chain,DINO,Coati,ink,coala} to create and maintain \emph{persistent state} on non-volatile memory (NVM).
These systems operate as the device approaches $V_{\mathit{off}}$, allowing systems to retain the application state across energy failures.
NVM operations, however, are extremely energy hungry and impose a significant energy overhead~\cite{maioli2021alfred}.

Experimental evidence exists that shows how intermittently-computing IoT devices are vulnerable to \emph{energy attacks}~\cite{mottola2023energy}, that is, energy harvesting can be used as an \emph{attack vector}.
Exerting simple control on how the ambient provisions energy allows an adversary to create situations of \emph{livelock}, \emph{priority inversion}, and \emph{denial of service}, \emph{without} requiring physical access to the devices.
The simplest scenario may consist, for example, in the attacker physically blocking the radiation arriving at a solar panel that powers the IoT device, eventually impeding forward progress, yet existing literature demonstrates more subtle setups~\cite{mottola2023energy}.

\fakepar{Contribution and road-map} We tackle the problem of \emph{mitigating} the impact of energy attacks on system performance.
One option is to design countermeasures that apply to specific attacks.
For example, when an energy attack is meant to cause a livelock, one may keep track of the number of restarts from a specific point in the code and suspend the operation for a given time period if this quantity exceeds a threshold.

We take a different stand here.
We develop a generic abstraction and support run-time, called Energy Attack Mitigation (\eam), that applies independent of the specific energy attack.
We consider a task-based multi-capacitor architecture, further described in \secref{sec:related}, given its use in battery-less IoT deployments~\cite{capybara,sensys20deployment, denby2023kodan}.
Using \eam, as illustrated in \secref{sec:design}, programmers specify different \emph{application profiles} that map task execution rates to different system states.
When an on-device attack detection system reports the occurrence of an energy attack~\cite{mottola2023energy}, \eam intelligently throttles task execution rates and accordingly orchestrates the charging of different capacitors, reducing energy consumption in an attempt to retain a minimum programmer-specified.
It resumes normal operation once the attack is (expected to be) over. EAM consists of the energy manager, task scheduler/peripheral control, and application manager functional blocks that are implemented in software and run on-board the IoT device MCU.

Our evaluation, reported in \secref{sec:eval}, uses a mixture of numerical simulations and emulation experiments to compare \eam against two state-of-the-art baselines, using \emph{real-world} energy traces. We evaluate the run-time energy consumption overhead associated with the on-board execution of the EAM solution. A single execution of EAM has negligible overhead, requiring less than 2 nJ energy and taking less than 1.5 $\mu$sec.  
The results indicate that \eam ensures the execution of 23.3\% additional application cycles compared to the baselines we consider and increases task schedulability by at least 21\%, while enabling a 34\%  higher component availability. 

\section{Background and Related Work}
\label{sec:related}

The security literature for \emph{battery-powered IoT} systems is extensive.
Resource-constraints generally make it difficult to apply standard security mechanisms~\cite{thakor2021lightweight}; further, battery-powered IoT devices enables peculiar attacks, for example, in an attempt to drain batteries~\cite{krentz2017countering,nguyen2019energy}.
To save energy, they feature low-power radios, which makes them vulnerable to denial of service attacks, for example, due to intentional jamming~\cite{kanwar2021jamsense}.
Specialized network stacks are also necessary for multi-hop operation, which are vulnerable to new kinds of attacks.
System characteristics motivate new techniques for attack detection and mitigation ranging from hardware-based solutions~\cite{portilla2010adaptable} to techniques employing machine learning~\cite{da2019internet,tahsien2020machine}.

Solutions for battery-powered IoT systems, unfortunately, falls short of expectations in the case of \emph{battery-less} IoT devices.
Energy constraints are way more severe compared to their battery-powered counterpart.
Limited form factors impose restrictions on the harvesting unit, limiting power supply to tens of $mW$~\cite{water-deployment-microbial-fuel-cell,pible,capacity-over-capacitance,ea_base}.
This creates a demand-supply gap.
Systems use tiny energy buffers, such as capacitors, to tame fluctuations of energy intake and for performing operations whose power consumption exceeds the maximum harvesting capabilities.
Moreover, the intermittent execution pattern adds a new dimension to the problem~\cite{awesome}, as periods of active operation are interspersed with periods for recharging energy buffers, while the rest of the system is quiescent.

\fakepar{Architectures} \figref{fig:arch} depicts the prevailing architecture for intermit\-tently-computing IoT devices, as seen in both available platforms~\cite{capacity-over-capacitance,flicker} and concrete deployments~\cite{water-deployment-microbial-fuel-cell,pible}.
A mixed-volatile MCU~\cite{msp430fr5969} relies on multiple capacitors~\cite{capybara,Hester2015} to tame fluctuations of energy intake and for performing operations whose power consumption exceeds the maximum harvesting capabilities.
The multiple capacitors allow the system to strike a better trade-off between charging times and available energy.
Smaller capacitors are the first to reach $V_{on}$; as this happens, tasks that consume little energy, such as probing low-power sensors, are immediately executed.
Bigger capacitors take longer to reach $V_{on}$; their energy is eventually consumed by energy-hungry tasks, such as radio operations.
A dedicated energy management module governs the charging of the multiple capacitors based on energy intake and task demands.

Intermittently-computing IoT devices employ techniques such as checkpointing~\cite{HarvOS,mementos,Hibernus,Hibernus++,chinchilla,QuickRecall,Samoyed,ratchet,dice} or task-based programming abstractions~\cite{chain,DINO,Coati,ink,coala,singhal2023} to deal with energy failures.
The former consist in replicating the application state on NVM, where it is retrieved back once the system resumes with sufficient energy.
The latter offer abstractions to define and manage persistent state, while taking care of data consistency in case of repeated executions of non-idempotent code~\cite{ratchet}.

We adopt a task-based programming model and a multi-capacitor architecture as their combination is demonstrated to be effective in real-world deployments~\cite{sensys20deployment,denby2023kodan,capybara}.
Crucially, they also offer ample opportunities to intelligently manage energy when the system is under attack, as illustrated in \secref{sec:design}.


\fakepar{Resource management} Orthogonal to the use of multi-capacitor architectures, several techniques exist to manage energy in battery-less IoT devices.
Dynamic voltage scaling (DVS) is one such technique that maintains the required performance level by adjusting the system supply voltage to a suitable minimum~\cite{T4}.
Dynamic frequency scaling (DFS), differently, meets the required performance by modulating the system clock frequency ~\cite{T1}.
Dynamic voltage and frequency scaling (DVFS) combines the two~\cite{T8,ahmed2020intermittent}; custom designs of DVFS exist especially for intermittently-computing devices.

Software-based maximum power point (MPP) tracking can achieve further energy gains.
It can adapt the power consumption of the system that is operating at an efficient operating voltage and maximizing forward application execution without adding any external tracking or control units~\cite{T9,T33}.
Custom MPP designs also exist for specific energy sources.
In the case of kinetic energy~\cite{T79}, for example, voltage-current characteristics of the energy transducer are used to find the optimal operating point, dynamically~\cite{T75}.

Energy scarcity also makes the real-time task scheduling difficult~\cite{islam2020, karimi2021,ea_base}.
Schedulability may be improved by dynamically scheduling the computational and energy-harvesting tasks~\cite{islam2020}.
This essentially means co-designing scheduling and energy management, which is demonstrate to be particularly effective in multi-capacitor architectures~\cite{islam2020,Hester2015}.
This is precisely what we do in this work, yet with a different goal compared to existing literature.
Instead of generally improving schedulability, we aim to reach a minimum programmer-specified task execution rate agains widely varying energy intakes, possibly affected by energy attacks.

\fakepar{Security} Compared to battery-powered IoT devices, the security issues in battery-less IoT devices drastically change, largely making the area uncharted territory.

The few existing solutions focus on securing persistent state.
Krishnan et al.~\cite{Krishnan2018} present an attack model for unsecured and cryptographically secured checkpoints.
Asad et al.~\cite{asad2020securing} present an experimental evaluation on the use of different encryption algorithms and ARM TrustZone.
Krishnan et al.~\cite{10.1145/3522748} build on this and propose a configurable checkpoint security setting that leverages application properties to reduce overhead. 
Ghodsi et al.~\cite{ghodsi2017optimal} use lightweight algorithms~\cite{borghoff2012prince} for securing checkpoints.
Valea et al.~\cite{valea2018si} propose a SECure Context Saving (SECCS) hardware module inside the MCU, whereas Khrishnan et al.~\cite{krishnan2019secure} apply Authenticated Encryption with Associated Data (AEAD) to protect checkpointing data.

Experimental evidence of energy attacks and corresponding vulnerabilities exists~\cite{mottola2023energy}, where machine learning is applied to develop a generic attack detection module.
This work tackles the complementary problem, that is, to mitigate the effects of an energy attack on system performance once the attack is detected.

\section{Design} \label{s:system} \label{sec:design}

We use three paradigmatic applications as running examples, described in \secref{sec:apps}, along with the programming model we use to encode their logic. 
\secref{sec:arch} describes the architecture of an \eam-enabled system and its run-time operation.

\subsection{Applications and Programming Model}\label{sec:apps}

\begin{figure}[tb]
\centering
\includegraphics[width=3.3in]{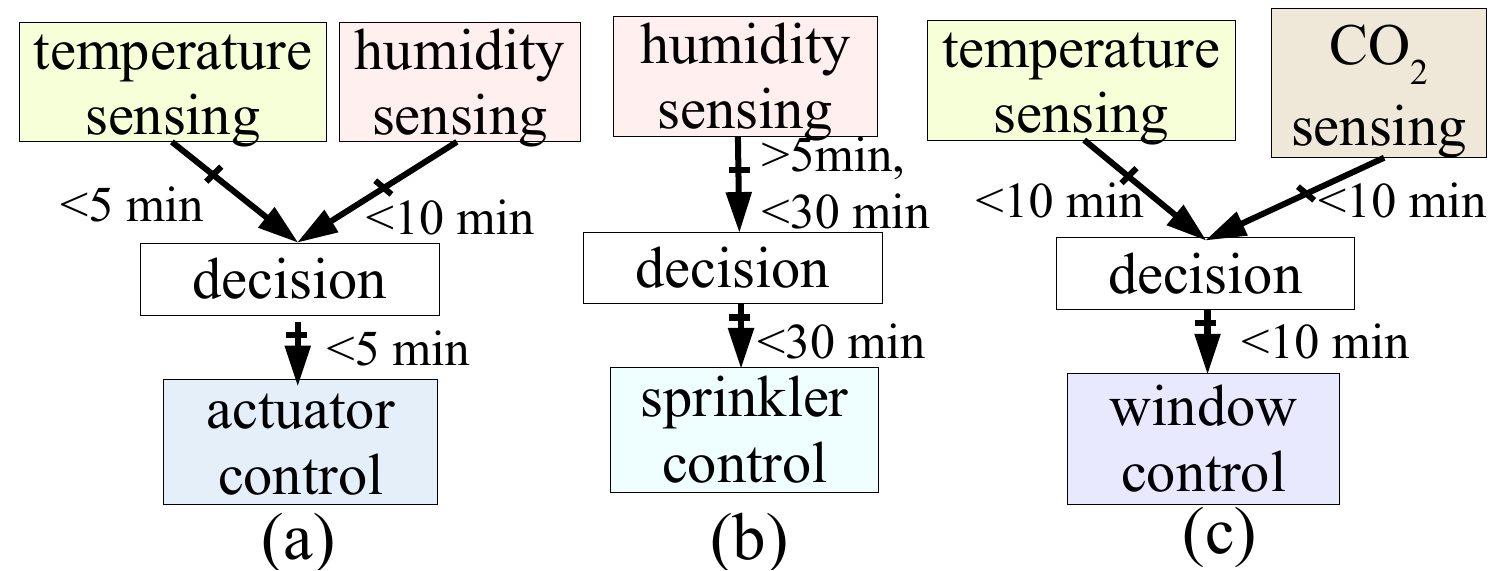}
\vspace{-0.15in}
\caption{Tasks in example applications: (a) HVAC system, (b) greenhouse irrigation system, and (c) controlled room ventilation. \capt{Tasks represent individual units of processing and run with transactional semantics. The exchange data through non-volatile data queues subject to timing requirements.}}
\vspace{-0.15in}
\label{fig:apps}
\end{figure}

We use a programming model that naturally maps to embedded sensing applications~\cite{amiri2017survey} and is based on well-established IoT programming concepts~\cite{Mottola2019,mottola2007enabling}. 
Using this model, we consider three example applications, graphically
illustrated in \figref{fig:apps}, corresponding to concrete deployments of battery-less IoT systems~\cite{flicker,sensys20deployment, water-deployment-microbial-fuel-cell}.

The application logic is split in individual tasks, shown as boxes in \figref{fig:apps}, which represent independent units of processing that execute asynchronously with respect to each other.
Tasks have inputs and outputs, represented in \figref{fig:apps} with links between the tasks, with attached timing requirements.
For example, task \emph{temperature sensing} in \figref{fig:apps}(a) must provide a new temperature reading to the \emph{decision} task at least once every five minutes.
Differently, task \emph{humidity sensing} in \figref{fig:apps}(b) must provide a new humidity reading to the \emph{decision} task at least once every 30 minutes, but no more frequently than once every five minutes~\cite{flicker,Hester2015,awesome}.

Tasks exchange data through non-volatile data queues, which offer a way to persist state in case of an energy failure.
Tasks run with transactional semantics; if they are interrupted by an energy failure before producing an output, no intermediate result is persisted and the task restarts from the beginning~\cite{dice,alpaca,chain,DINO,ink}.
When a data item flows through the entire task pipeline from start to end, we say the application has \emph{completed an execution}. 
The application execution rate, that is, the number of times an application completes an execution in the unit of time, is the primary performance metric we discuss in \secref{sec:eval}.

\subsection{Architecture and Run-time} \label{sec:arch}

\begin{figure}[tb]
\centering
\includegraphics[width=3.5in]{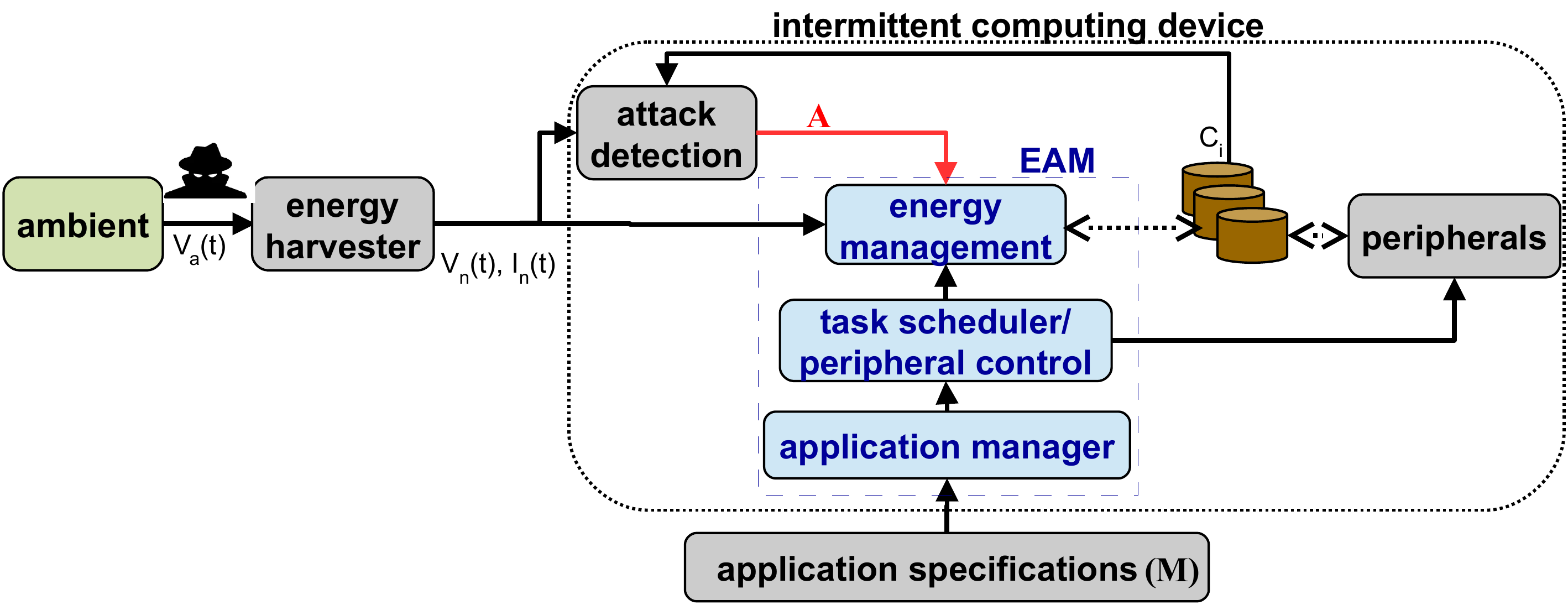}
\vspace{-0.15in}
\caption{Hardware/software device architecture when using \eam. \capt{\eam takes as input the output of the attack detection module, the $V_n(t), I_n(t)$ signals from the energy harvester, the current capacitor charge, and the application specification $\mathbf{M}$.}}
\vspace{-0.15in}
\label{fig:arch}
\end{figure}

\begin{figure}[tb]
\small
\begin{center}
\begin{tabular}{|p{0.95in}|p{2.2in}|} 
 \hline
 \bf{Parameter} & \bf{Value} \\ \hline
{\bf Supported profiles} & Short attack (SA), long attack (LA), normal (NML), low power (LP), critical (CTL)\\ \hline
{\bf Task set} & {\{Tasks\}}\\ \hline
{\bf Task rate} & {Profile: \{Tasks (Rate)\}}\\ \hline
{\bf Task dependency} & {\{Finish-to-start Task Lists\}}\\ \hline
\end{tabular}
\end{center}
\vspace{-0.15in}
\caption{Application specification $\mathbf{M}$. \capt{It includes information on the profiles supported by an application, the task set, their rate requirements, and their dependencies.}}
\vspace{-0.15in}
\label{t:met}
\label{fig:informationM}
\end{figure}

\begin{figure}[tb]
\small
\begin{center}
\begin{tabular}{ |c|c| } 
 \hline
 \bf{Parameter} & \bf{Symbol} \\ \hline
 Attack ongoing & $a_o$\\ \hline
 Attack ongoing accuracy & $a_{oa}$ \\ \hline
 Elapsed time & $a_{et}$  \\\hline
 Remaining time & $a_{rt}$\\  \hline
 \end{tabular}
\end{center}
\vspace{-0.15in}
\caption{Energy attack parameters in information set $\mathbf{A}$. \capt{The information returned by the attack detection module as used by \eam to adapt the application behavior during an attack.}}
\vspace{-0.15in}
\label{fig:informationA}
\label{t:ea_param}
\end{figure}

\begin{figure}[tb]
\small
\begin{center}
\begin{tabular}{ |c|l| } 
 \hline
 \bf{Symbol} & \bf{Definition} \\ \hline
$\rho$&Application profile\\ \hline
 $\alpha$ & Maximum $a_{rt}$ corresponding to SA \\ \hline
 $E,E_i$&Available energy and energy stored in buffer $i$\\ \hline
 $\Omega_0,\Omega_1$ & Maximum $E$ for CTL and LP profiles, resp. \\ \hline
 $\tau,\mathbf{T},\tau_e$&Task, task set, and current execution task\\ \hline
\!\!\!$\mathbf{R}(\tau),\mathbf{S}(\tau)$\!\!\!& Task-rate and Task-state of $\tau$, resp. \\\hline
\!\!\!$\epsilon(\tau), \mathcal{B}(\tau)$\!\!\!& Energy required and energy-buffer used to execute $\tau$\\  \hline
$\lambda_\tau$ &\!\!\!Energy proportion stored in $\mathcal{B}(\tau)$ for blocked, suspended $\tau$\!\!\!\\ \hline
\!\!\!$\Lambda_\tau$,$\Lambda_\tau>\lambda_\tau$\!\!\!& Energy proportion stored in $\mathcal{B}(\tau)$ for ready, running $\tau$\\ \hline

 \end{tabular}
\end{center}
\vspace{-0.15in}
\caption{Notation. \capt{The EAM parameter symbols and corresponding definitions are summarized here.}}
\vspace{-0.15in}
\label{fig:notation}
\label{t:notation}
\end{figure}

\figref{fig:arch} illustrates the hardware/software architecture of an
\eam-enabled system.
\eam takes as input a specification $\mathbf{M}$ of the application behavior, illustrated in \figref{fig:informationM}.
This specification includes information on what \emph{system profile} is supported by the application among five possible options we discuss next.
It also specifies the application's task set, along with their minimum and maximum required execution rates for each supported profile and their dependencies.
The latter is essentially an encoding of the information graphically represented in \figref{fig:apps}.  

The five available system profiles allow programmers to tune an application's behavior depending on the situation.
Profile \emph{normal} (NML) indicates regular operation. Profile \emph{low-power} (LP) represents a general situation of energy scarcity, yet the application may still continue the execution, whereas profile \emph{critical} (CTL) represents a situation where the system may fail to achieve progress due to extreme energy scarcity.
None of these situations, however, necessarily indicates that an adversary is manipulating the energy supplies.
These situations are represented by profile \emph{short attack (SA)} and \emph{long attack (LA)}, depending on the expected attack duration.
We indicate with $\rho$ what system profile is currently active.


The adversary manipulates the energy signal $V_a(t)$ at time $t$ and affects the voltage and current $V_n(t), I_n(t)$ supplied to the intermittently-computing IoT device.
The attack detection module, shown in \figref{fig:arch} may realize an attack is occurring and alert \eam by passing an information set $\mathbf{A}$ about the attack, along with the manipulated $V_n(t), I_n(t)$ signals.
The information we expect to find in $\mathbf{A}$ is in \figref{fig:informationA}.
Quantity $a_o$ is a binary flag indicating whether an attack is ongoing, whereas quantity $a_{oa}$ indicates how reliable is the information provided by the attack detector.
Quantity $a_{et}$ and $a_{rt}$ respectively indicate when the attack detector thinks the attack started, and for how long it is expected to continue; these quantities are useful to determine whether profile SA or LA better represent the current situation.
Existing attack detectors can provide this information accurately~\cite{mottola2023energy}. We consider that the harvester receives no energy during the the long and short energy attack, that is, during an ongoing energy attack ($a_o=1$) at time $t'$, $V_n(t')=0$. 

In \figref{fig:arch}, boxes in light blue are the modules we add to a regular configuration.
We describe their functioning next.

\begin{figure}[tb]
\centering
\includegraphics[width=3.5in]{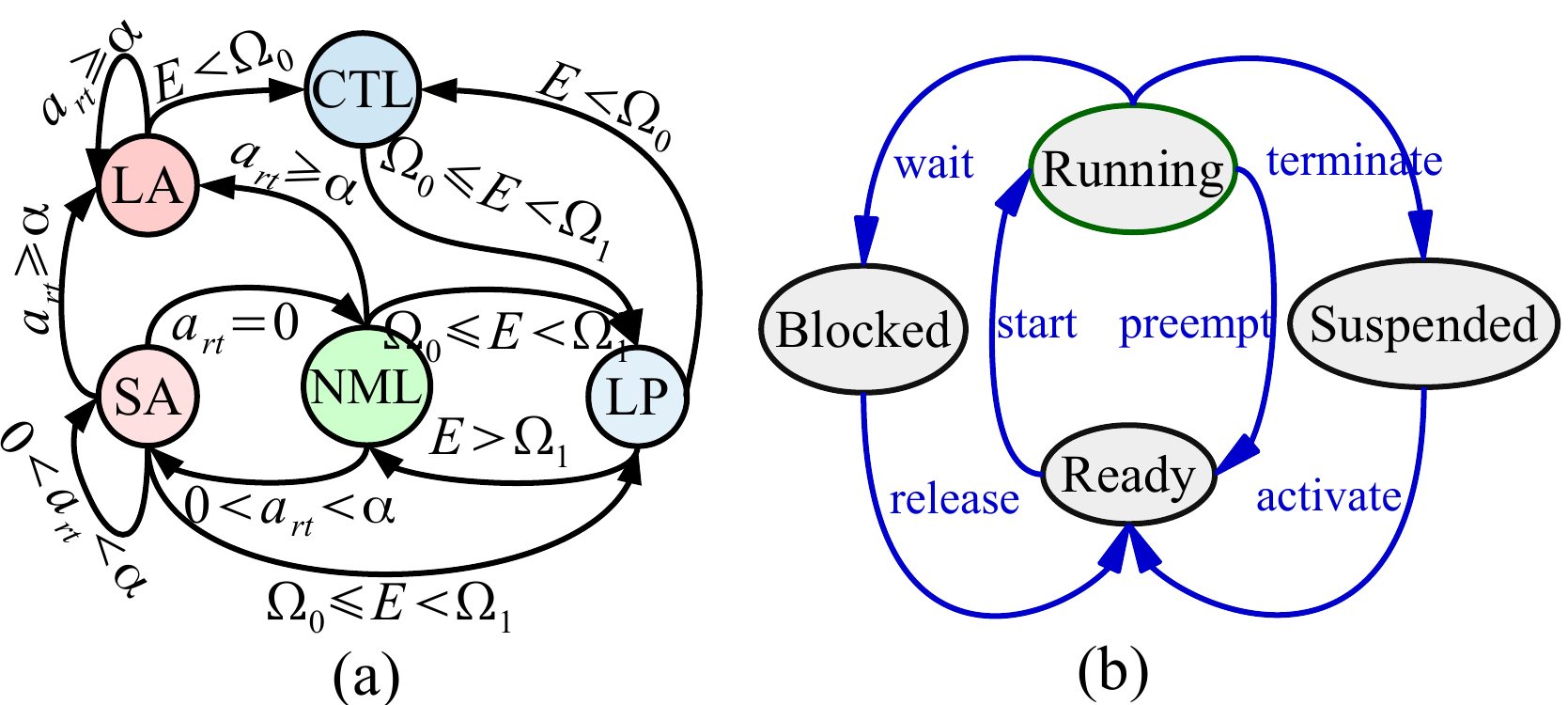}
\vspace{-0.27in}
\caption{Transition diagrams for: (a) system profile and (b) task state. \capt{System profiles model configurations that \eam enforces depending on the situation, for example, in case of an energy attack.}}
\vspace{-0.2in}
\label{f:ex2}
\end{figure}



\fakepar{Application manager} The \emph{application manager} module governs the application execution depending on the active system profile, based on the application specification $\mathbf{M}$.
It does so by periodically probing the energy buffer to estimate the available energy $E$ and by monitoring the information set $\mathbf{A}$ about possible attacks.

\figref{f:ex2}(a) shows the transition diagram for system profile.
The application manager selects the system profile based on available energy $E$ and the energy attack parameter $a_{rt}$.
When an attack is ongoing, that is, as long as $a_o=1$, if the remaining time of the attack is greater than a configuration parameter $\alpha$, that is, $a_{rt}\geq \alpha$, then the LA profile is selected; or else the SA is selected. 
The energy level thresholds $\Omega_0$ and $\Omega_1$, $\Omega_1>\Omega_0$, are used to select between the NML, LP, and CTL profiles.
When the attack is not ongoing, that is, as long as $a_o=0$, if $E>\Omega_1$ then the application manager selects the NML profile; or else if  $E<\Omega_0$ then the CTL profile is selected.
If none of these applies, the LP profile is selected. The system state, including energy level and energy attack information, is periodically updated and profile selection happens based on the system current state.

\fakepar{Energy manager} We consider a multi-capacitor energy storage architecture~\cite{Hester2015}, which allows the system to strike an efficient trade-off between charging times and energy efficiency.
The lightweight tasks consume energy from the smaller capacitors that charge more quickly, whereas more energy-intensive tasks are executed using the energy stored in the larger capacitors~\cite{capybara}.

We consider a parallel resistor-capacitor circuit with the equivalent storage capacitor, $C_i$ in parallel to the resistor equivalent to the rest of the circuit, $R_p$. Hence the voltage of the capacitor, $C_i$ at time instance $t$, during charging, is given as:
\begin{equation}
    V_i(t)=\sqrt{P(t)\cdot R_p - e^\frac{-2t}{C_i\cdot R_p}\cdot \left( P\cdot R_p - V_0^2\right)}
\end{equation}
where $V_0$ is the capacitor voltage at $t=0$ and $P(t)$ is the power from the energy source at time $t$. 
The energy buffer, $\mathcal{B}(\tau)\in \mathbb{N}$, $\mathcal{B}(\tau)\leq m$ is the capacitor that supports execution of task $\tau$. 

The \emph{energy manager} module in \figref{fig:arch} directs the incoming energy towards different capacitors depending on the task energy requirements in the different system profiles.
It stores energy proportionally in the capacitors that sustain the operation of tasks that are ready to execute, that is, they have data in their input queues, and are closer to the deadline indicated in the application specification.
This means that the energy manager dynamically changes the amount of energy going towards different capacitors depending on the instantaneous system needs.
This is achieved by periodically checking the state of all tasks, the corresponding timing requirements given the active system profile, and the current energy content of all capacitors.

\fakepar{Task scheduler and peripheral control}
We \emph{co-design} task scheduling with energy management.
Tasks may be in one of the four states of \figref{f:ex2}(b).
Tasks are ready when there is data in their incoming queues.
One of them is selected for execution using a form of real-time periodic task scheduling (RTS)~\cite{karimi2021}, subject to sufficient energy availability in the corresponding capacitor.
This means that a task that is ready to execute, but has no sufficient energy to complete, is deferred until sufficient energy is available.

Tasks are set to be in a blocked state if they depend  on another task that is yet to be executed, that is, they consume its output.
Suspended tasks are those waiting for low-level I/O operations to complete, for example, obtaining a sensor reading, or are waiting for sufficient energy to start the execution.

\begin{figure}[tb]
  \centering
    \includegraphics[width=.99\linewidth]{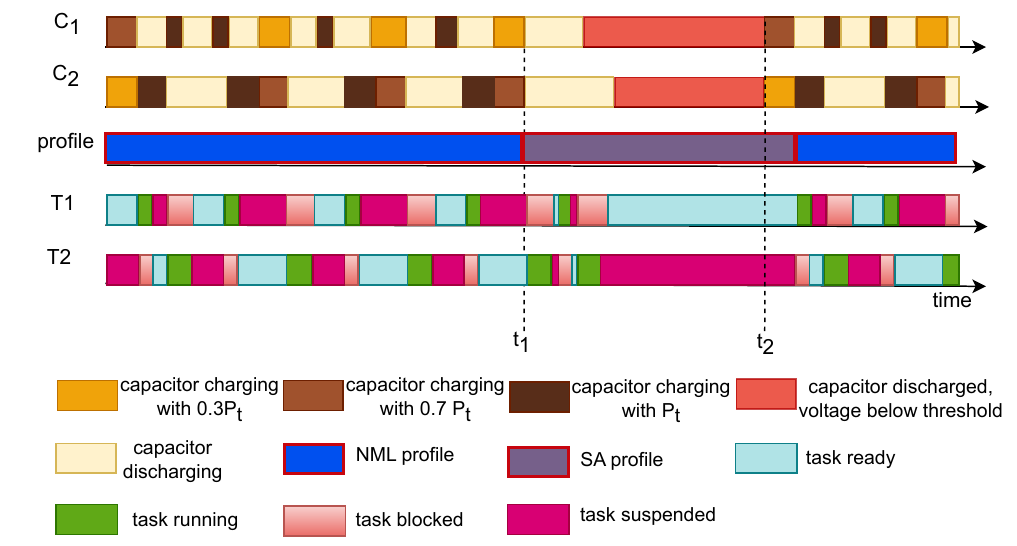}
    \vspace{-0.15in}
   \caption{Example \eam execution. \capt{A short energy attack is detected at $t_1$. The system transitions to profile SA where only T1 is scheduled for execution as all incoming energy is directed to $C_1$. T2 is suspended in the mean time and its capacitor does not charge until the attack is over and the system transitions back to NLM.}}
   \vspace{-0.2in}
   \label{fig:exampleEAM}
 \end{figure}

\fakepar{Example} \figref{fig:exampleEAM} shows a timing diagram for an example application that has two tasks, $T_1$ (light weight) and $T_2$ (energy-intensive), that are executed using the energy stored in capacitors $C_1$ and $C_2$, respectively. Here, $C_1<C_2$, thus $C_1$ has shorter charging and discharging times as compared to $C_2$. Task $T_2$ depends on the input from $T_1$ and can only execute after completion of $T_1$. The capacitors are charged proportionally with input power based on the state of the tasks that each supports. A higher proportion of power is used to charge the capacitor that must support a ready task.  

Initially, the profile is set to NML. A short attack happens from $t_1$ to $t_2$ and the energy available for charging the capacitors greatly reduces during this time. When the energy attack starts at $t_1$, the profile is set to SA. During the energy attack,  the tasks are executed as many times as possible while maintaining dependency using the remaining energy in the capacitors until they both fall below the threshold. When the energy attack ends at time $t_2$, the capacitors start charging proportionally based on the task state. As soon as the voltage in $C_1$ reaches $V_{on}$, \eam is executed and the profile is set back to NML and the periodic execution of the tasks resumes.

\fakepar{Energy attack mitigation algorithm}
Algorithm \ref{eamfh} shows EAM's operation through psuedocode. The inputs include the application specification $\mathbf{M}$ and energy attack parameters $\mathbf{A}$.  
The outputs include the system profile $\rho$, task states $\mathbf{S}$, and energy stored in capacitors $E_i, 1\leq i\leq m$. EAM includes three core functionality, operating as follows.
\begin{itemize}
\item Function \texttt{AppTaskManager} implements the \emph{application manager}. It selects the system profile $\rho$ based on the available energy and energy attack parameters $a_o$ and $a_{rt}$. Thereafter, it creates the task set $\mathbf{T}$ and assigns the task rate $\mathbf{R}(\tau), \forall \tau\in\mathbf{T}$ using the profile-specific application specification ($\mathbf{M}$). 
\item Function \texttt{TaskScheduler} implements \emph{task scheduling and peripheral control}. 
It sets the task state $\mathbf{S}(\tau)$ as ready if there is sufficient energy in the buffer $\mathcal{B}(\tau)$ to support the execution of task $\tau$. 
In the event of an energy attack, a task state is set to ready if the remaining time of the attack is estimated to be greater than the task execution time, that is, $a_{rt}>1/\mathbf{R}(\tau)$ while the capacitor supporting its execution has sufficient energy. Otherwise the task state is set to blocked. A ready task is selected to be the current execution task $\tau_e$ when its supporting capacitor has sufficient stored energy. 
\item Function \texttt{FederatedHarvesting} implements the \emph{energy manager}. It proportionally stores the harvested energy in separate capacitors based on the state of the task it supports. If the task $\tau$ is ready then a higher proportion of energy $\Lambda_\tau$ is stored in the corresponding buffer, else a lower proportion of energy $\lambda_\tau$ is stored.
\end{itemize}

\begin{algorithm}[!htb]
\small
\caption{ EAM: Energy-attack Mitigation}
\label{eamfh}
\begin{algorithmic}
\State \textbf{Input:} $\mathbf{A}$, $\mathbf{M}$\\
 \SetKwFunction{FMain}{Main}
  \SetKwFunction{FSum}{\!\!}
  \SetKwFunction{FSub}{\!\!}
  \SetKwProg{Fn}{AppTaskManager}{:}{}
  \Fn{\FSum{$\mathbf{A}$,$\mathbf{M}$}}{
\begin{enumerate}
 \item [1)] {\bf Select profile $\rho$:} 
  \end{enumerate}
 ${}$\hspace{1em} \If{$a_o=1$ }
  {
  \lIf{$a_{rt}>\alpha$}{
  $\rho=$ LA}
   \lElse{
   $\rho=$ SA}
   }
   \Else{
    \lIf{$\omega_o<E\leq \omega_1$}{
   $\rho=$ LP}
   \lElseIf{$E<\omega_0$}{$\rho=$ CTL}
   \lElse{$\rho=$ NML}
      }
  
\begin{enumerate}
\item [2)] {\bf Select task set:} \newline Include task $\tau$ in task set $\mathbf{T}$ if $\mathbf{M}(\tau,\rho)>0$
\end{enumerate}
  \begin{enumerate}
\item [3)] {\bf Assign task rate:} $\mathbf{R}(\tau)=\mathbf{M}(\tau,\rho)$
\end{enumerate}
\KwRet $\rho$, $\mathbf{T}$, $\mathbf{R}$
}
\noindent\SetKwProg{Fn}{TaskScheduler}{:}{}
  \Fn{\FSub{$\rho$, $\mathbf{T}$, $\mathbf{R}$, $\mathbf{A}$}}{
\begin{enumerate}
     \item [1)] {\bf Set task state (when there is no energy attack):}
\end{enumerate}
  \For{each task $\tau\in \mathbf{T}$}{
   ${}$\hspace{-2em}\lIf{(($E_{\mathcal{B}(\tau)}\geq\varepsilon(\tau)$) and ($\tau=1$\,or\,$\mathbf{S}(\tau-1)=$`running'))}{{\bf set} $\mathbf{S}(\tau)=$`ready'}
  ${}$\hspace{-2em}\lElse{{\bf set} $\mathbf{S}(\tau)=$`blocked'}}
  \begin{enumerate}
     \item [2)] {\bf Defer/ execute task in event of energy attack, $a_o=1$:}
\end{enumerate}
${}$\hspace{0.2em}\For{ each task $\tau\in \mathbf{T}$}{
  \lIf{ $a_{rt}>(1/R(\tau))$ and $E_{\mathcal{B}(\tau)}>\varepsilon(\tau)$}{\newline${}$\hspace{2.5em}{\bf set} $\mathbf{S}(\tau)=$`ready'}\lElse{{\bf set} $\mathbf{S}(\tau)=$`blocked'}}
    \begin{enumerate}
     \item [3)] {\bf Execution task $\tau_e$ from the task set $\mathbf{T}$:}
\end{enumerate}
  ${}$\hspace{0.2em}\For{each task $\tau\in \mathbf{T}$}{
  \hspace{-1.5em}\If{$\mathbf{S}(\tau)=$`ready' and $E_{\mathcal{B}(\tau)}\geq\varepsilon(\tau)$}{ ${}$\hspace{0.5em}{\bf Set} $\tau_e=\tau$\\${}$\hspace{2.5em}{\bf Set} $\mathbf{S}(\tau)=$`running'\\${}$\hspace{2.5em}{\bf Remove} $\tau$ from $\mathbf{T}$}}
  \KwRet $\mathbf{S}$, $\tau_e$
  }
\noindent\SetKwProg{Fn}{FederatedHarvesting}{:}{}
  \Fn{\FSub{$\rho$, $\mathbf{T}$, $\mathbf{R}$, $\mathbf{S}$}}{
  \begin{enumerate}
     \item [1)] {\bf Proportion of harvested energy to buffers}
\end{enumerate}
 ${}$\hspace{1em}\For{each buffer $i=1$ to $m$}{\For{each $\tau \in\mathbf{T}$}{
\lIf{$\mathbf{S}(\tau)=$`ready' or `running'}{\newline${}$\hspace{2.8em}{\bf Set} $E_{\mathcal{B}(\tau)}=\Lambda_\tau\cdot E$}
  \lElse{{\bf Set} $E_{\mathcal{B}(\tau)}=\lambda_\tau\cdot E$}}}
  \KwRet $E_i(n)$
  }

\noindent\textbf{Output:} $\rho, \mathbf{S}, E_i, 1\leq i \leq m$
\end{algorithmic}
\end{algorithm}
\vspace{-0.1in}

\section{Evaluation}
\label{sec:eval}
We illustrate the experimental setup in \secref{sec:setting} and split the discussion of the results in three parts.
In \secref{sec:rate}, we investigate the improvement enabled by \eam in the \emph{application execution rate}, that is, the number of times an application manages to complete the data pipeline.
Next, in \secref{sec:scheduling}, we study the \emph{task scheduling} behavior to gain a deeper understanding of how \eam enables the improvements in application execution rate, and conclude the discussion with an analysis of the \emph{run-time energy overhead} imposed by \eam.

\subsection{Setting}
\label{sec:setting}

We use the application in \figref{fig:apps}(a) to experimentally measure the performance of \eam and of the baselines.
This is the most challenging case in \figref{fig:apps}, especially in terms of task execution rates. 
We describe next the metrics we use to evaluate \eam performance, along with the baselines we compare with  and the tools we use.

\fakepar{Metrics} We consider five performance metrics
\begin{itemize}
\item The \emph{application execution rate} is the number of times per time unit an application completes the data pipeline, while meeting all task dependencies. 
This is a function of the individual task execution rates, indicated in the application specification $\mathbf{M}$ depending on the current system profile, and of the way the system manages available resources. 
    \item The \emph{task schedulability} measures the percentage of time that tasks in the system have sufficient resources to execute: in our specific setting, this means that the capacitor(s) in the system store sufficient energy for the task to execute and produce the corresponding output. 
    \item The \emph{component availability} is the percentage of time that the MCU (for computation) or the attached peripherals (for sensing, wireless transmissions, or actuation) have sufficient resources to execute. This metric essentially represents the hardware counterpart of the task schedulability figure. 
    \item The \emph{time and energy overhead} is the time or energy spent for running the given scheduling technique, which essentially measures how much time or energy  is subtracted from application processing to manage system resources.
    \end{itemize}

The application execution rate represents the primary metric, because it measures the level of service provided to end users.
Ideally, the scheduler should \emph{at least} meet the application specification $\mathbf{M}$, while any other improvement is welcome as it indicates the system operates above the minimum requirement.
The other metrics, but the run-time energy overhead, are instrumental to understand the trends we observe in application execution rate.
For example, a higher execution rate should be enabled by higher task schedulability and/or component availability.
The run-time energy overhead is the cost we pay for managing system resources in a given way.
Measuring this quantity is necessary to weigh the possible improvements in application execution rate against the needed energy cost.

\begin{figure}[tb]
\centering
\includegraphics[width=.99\linewidth]{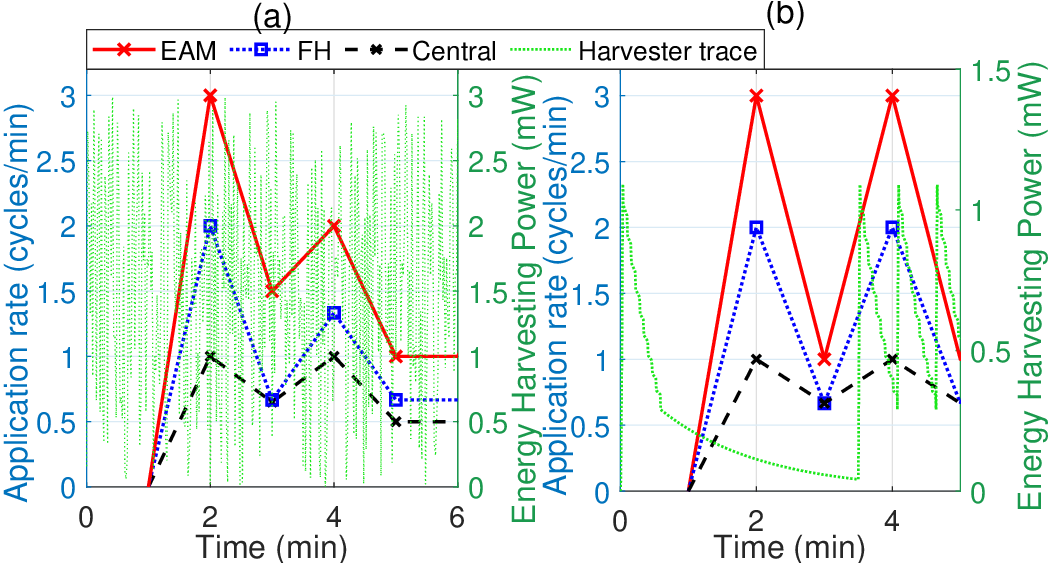}
\vspace{-0.15in}
\caption{Application execution rate for (a) outdoor and (b) indoor traces using \eam, FH, or Central.}
\vspace{-0.15in}
\label{f:ar}
\end{figure}

\fakepar{Baselines and tools} We compare \eam with federated energy harvesting (FH)~\cite{flicker} and a centralized architecture (Central)~\cite{Hester2015}.
Both use the same real-time energy-aware periodic task scheduling (RTS)~\cite{karimi2021,ea_base} as \eam.
FH is a multi-capacitor architecture that statically directs incoming energy to the multiple capacitors, in a way proportional to capacitor size.
This means it does not take into account possibly varying task execution rates.
FH uses the same number and type of capacitors we use in \eam.
Comparing with FH allows us to measure the gains enabled by making the scheduler aware of changing application requirements, based on the currently active system profile.
The Central baseline, instead, allows us to study the benefits brought by combining a multi-capacitor architecture with application-aware energy management and scheduling.

To obtain quantitative results, we use a combination of numerical simulations and emulation.
We use Matlab to evaluate the performance of \eam and the baselines in application execution rate, task schedulability, component availability, and component availability latency.
We implement accurate models of energy harvesting and energy consumption, based on the operation of a staple MSP430-based intermittent computing platform~\cite{flicker}.
We compute the run-time energy overhead using MSPSim, which provides time-accurate emulation of binary code for the MSP430 platform, considering a 3V capacitor supply as in existing literature~\cite{epic}.

\begin{figure*}[tb]
\centering
\includegraphics[width=.99\linewidth]{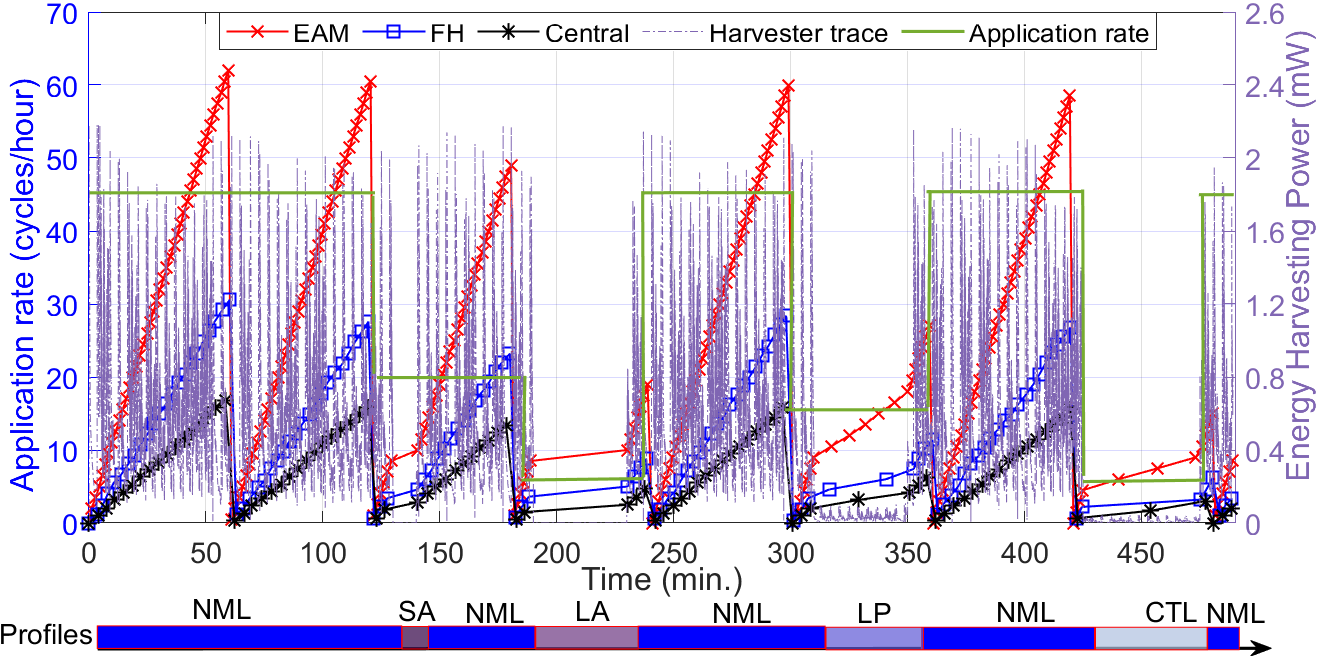}
\vspace*{-0.15in}
\caption{Application execution rate and active system profile.}
\vspace*{-0.15in}
\label{f:trace_ext}
\end{figure*}

We feed the numerical models with \emph{real-world} energy traces.
The traces are time series obtained from the actual output voltage across a 30k$\Omega$ resistor in a solar harvesting system using a monocrystalline high-efficiency solar panel that is partly stationary and partly moving, and operates indoor or outdoors~\cite{epic}. 
To investigate the performance during an attack, we also emulate different types of energy attacks, either short- or long-term, by completely zeroing the input voltage trace for a variable time period~\cite{mottola2023energy}.

\subsection{Application Execution Rate}
\label{sec:rate}

\figref{f:ar} 
 shows an example trace across a few minutes of execution demonstrating how the application execution rate enabled by \eam is systematically higher than the baselines we consider.
This observation applies to both traces.
The plot also demonstrates how \eam can capture energy fluctuations more effectively than the baselines.
For example, at 2 and 4 minutes in \figref{f:ar}(a), the system ends up with an excess of energy and the application execution rate enabled by all solutions accordingly increases.
However, the relative improvement is much higher for \eam than for the baselines. 

\figref{f:trace_ext} 
 provides a closer look at the system behavior in case of the scenarios that correspond to the supported application profiles including a short and a long energy attack.
The application execution rate enabled by \eam is higher than the baselines in the absence of an energy attack.
At time 11175 
sec (186.25 min) a short attack begins and \eam maintains a higher application execution rate, resuming normal operation right after the attack is over and maintaining better performance than the baselines throughout.
The same observation applies at time 11268 sec 
(187.81 min) when a long attack occurs.

\begin{figure}[tb]
\centering
\includegraphics[width=.99\linewidth]{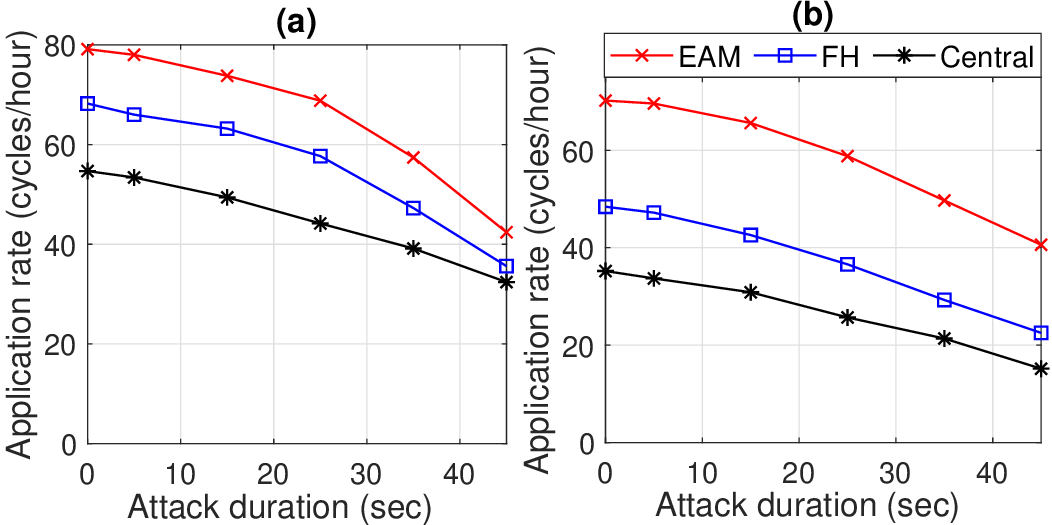}
\vspace{-0.15in}
\caption{Application execution rate in case of an energy attack for (a) outdoor and (b) indoor traces.}
\vspace{-0.2in}
\label{f:ea_2}
\end{figure}

Most importantly, \figref{f:ea_2} plots the application execution rate when we emulate an energy attack of variable duration at a random point in the energy trace, which is the same for \eam and the baselines.
As expected, the performance degrades with longer energy attacks, in that the energy budget available for system operation accordingly reduces.
However, \eam shows better and more robust performance: the gains over the baselines are retained regardless of the duration of the energy attack.
As before, this observation applies to both traces, demonstrating the generality of the conclusion.

\begin{figure}[tb]
\centering
\includegraphics[width=.99\linewidth]{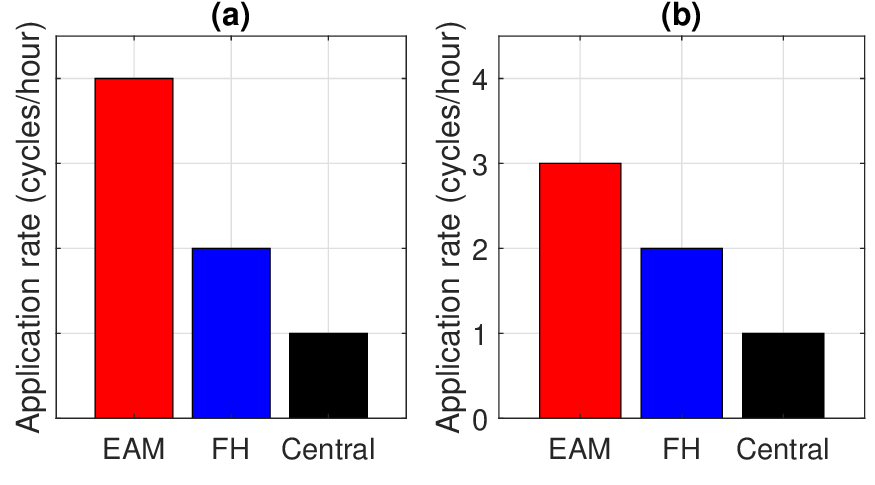}
\vspace{-0.15in}
\caption{Application execution rate when an attack commences and all solutions have the same energy budget for (a) outdoor and (b) indoor traces.}
\vspace{-0.2in}
\label{f:ea_3}
\end{figure}
\begin{figure}[tb]
\centering
\includegraphics[width=.99\linewidth]{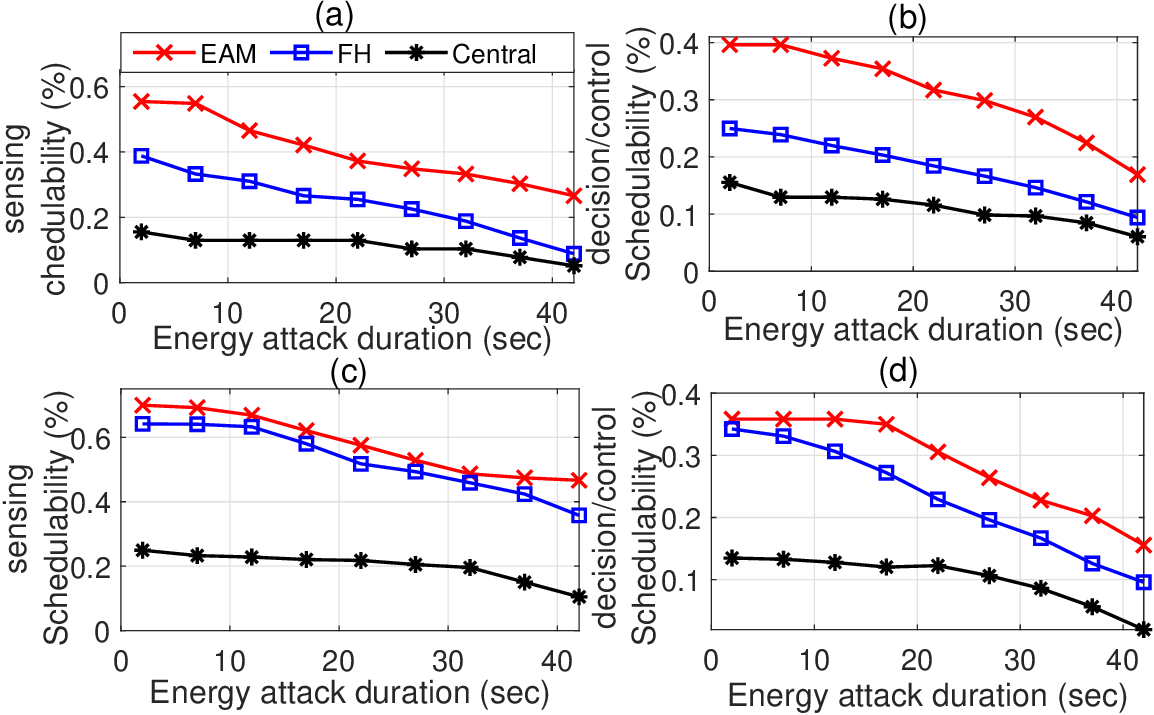}
\vspace{-0.15in}
\caption{Task schedulability depending with variable energy attack duration for (a-b) outdoor and (c-d) indoor traces.}
\vspace{-0.15in}
\label{f:ea_1}
\end{figure}

As \figref{f:ar} and \figref{f:trace_ext} demonstrate that \eam performs better than the baselines already in the absence of an attack, one may conjecture that the better resilience to energy attacks shown in \figref{f:ea_2} is not due to \eam's design, but comes from an energy excess gained beforehand that is eventually spent during the attack.
To investigate this specific aspect, we run experiments where both \eam and the baselines are given the same energy budget right before the attack commences, hence nullifying any energy excess that \eam may accumulate beforehand.
\figref{f:ea_3} 
shows the results.
Even when \eam is given the same initial energy budget than the baselines and an attack commences, \eam enables up to twice the application execution rate compared to the baselines.


\subsection{Scheduling}
\label{sec:scheduling}

The better performance in application execution rate comes from better resource management, which leads to more efficient task scheduling.
\figref{f:ea_1} 
 depicts the results we obtain in task schedulability, as a function of the type of task being scheduled, the attack duration, and the energy trace.
Regardless of the setting, \eam constantly ensures higher schedulability for both types of tasks and for both traces.
The result is thus arguably general.

\begin{figure}[tb]
\centering
\includegraphics[width=.99\linewidth]{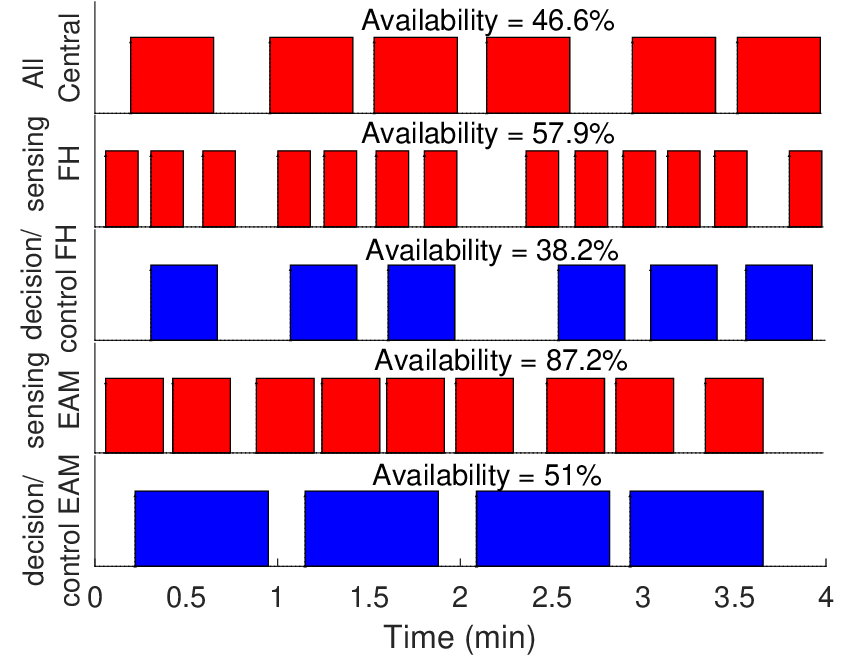}
\vspace{-0.15in}
\caption{Component availability using outdoor trace.}
\vspace{-0.15in}
\label{f:energy_data1}
\end{figure}

\begin{figure}[tb]
\centering
\includegraphics[width=.99\linewidth]{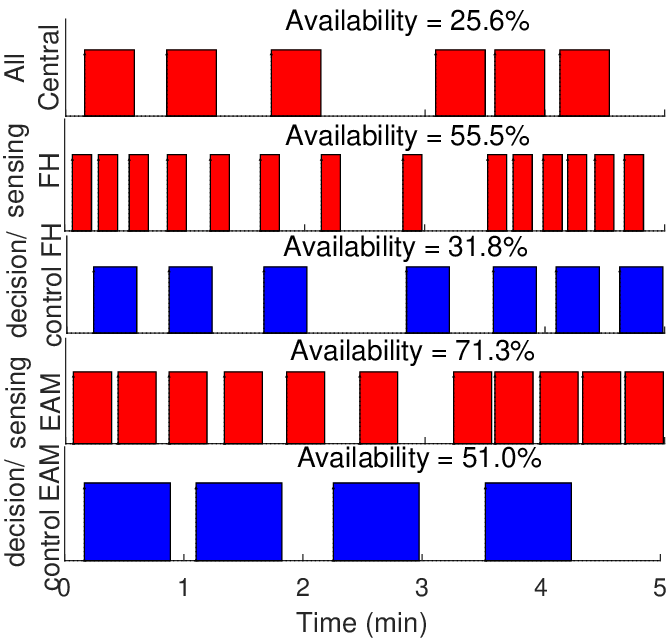}
\vspace{-0.15in}
\caption{Component availability using indoor trace.}
\vspace{-0.15in}
\label{f:energy_data2}
\end{figure}

\figref{f:energy_data1} and \figref{f:energy_data2} report the component availability of \eam and the baselines in the first few minutes of a sample execution and overall, for either energy trace.
Using outdoor trace, we observe that FH yields a higher availability for the MCU than Central, yet a lower availability for the peripherals necessary for actuation.
The situation is different in indoor trace, where FH improves the availability of both compared to Central.
Regardless of the Trace, \eam yields higher component availability than any other baseline.





We use MSPSim to ascertain the energy consumption overhead due to \eam, compared to the energy cost of running the actual application logic.
A single execution of \eam requires 1.781 nJ and takes 1.237  $\mu$sec.
The energy cost and execution time of a single application execution is reported in \figref{t:mspsim_results} depending on the single task \cite{msp430f2618-datasheet,msp430fr5969,ea_base}. 
The execution time  and energy consumption of \eam are orders of magnitude smaller than any application task.
The overhead is thus negligible.

\begin{figure}[tb]
\small
\begin{center}
\begin{tabular}{ |p{0.5in}|c|c| } 
 \hline
 \bf{Task} & \bf{Duration}&\bf{Energy} \\ \hline
Sensing & 12.030 msec & 19.066 $\mu$J\\ \hline
 Decision & 10.182 msec & 15.731 $\mu$J\\ \hline
 Control & 60.150 msec & 92.931 $\mu$J \\\hline
 \end{tabular}
 \vspace{-0.15in}
\caption{Energy cost of application tasks.}
\vspace{-0.15in}
\label{t:mspsim_results}
\end{center}
\end{figure}

\section{Conclusion} \label{sec:conclusions}

We presented \eam, an application-aware energy attack
mitigation system.
\eam tunes the task execution rates and steers the charging of multiple capacitors to meet programmer-provided application requirements, specified  as a function of system profiles.
We illustrated the co-design of scheduling and energy management in \eam and quantified its performance against two baselines.
We experimentally demonstrated that \eam produces higher application execution rates before, during, and after energy attacks, compared to the baselines we consider.
It ensures the execution of 23.3\% additional application cycles compared to the baselines we consider and increases task schedulability by at least 21\%, while enabling a 34\%  higher component availability.

\section*{Acknowledgement}
This work has been financially supported by the Swedish Foundation for
Strategic Research.



\setcitestyle{numbers,sort,compress}
\balance
 \bibliographystyle{ACM-Reference-Format}
 \bibliography{final,biblio,bibliography}
\end{document}